\documentstyle{article}
\date{}             
\begin{document}
\title{The suppression of $\pi NN(1535)$ coupling in QCD }
\author{Shi-Lin Zhu\\
Institute of Theoretical Physics, Academia Sinica \\ 
P.O.BOX 2735, Beijing 100080, P.R.China}
\maketitle
\begin{center}
\begin{minipage}{120mm}
\begin{center}{\bf Abstract}\end{center}
{\large
The light cone QCD sum rules are employed to calculate 
the $\pi NN(1535)$ coupling $g_{\pi NN^\ast} $.
We study the two point correlation function of two nucleon currents 
sandwiched between the vacuum and the pion state. 
The contribution from the excited states and the continuum is subtracted 
cleanly through the double Borel transform with respect to the nucleon 
and $N(1535)$ momenta, $p_1^2$, $p_2^2=(p-q)^2$.
Our calculation shows that the $\pi NN(1535)$ coupling is strongly 
suppressed.

\vskip 0.5 true cm
PACS Indices: 13.75.Gx; 14.20.Gk; 14.40.Aq; 13.75.Cs; 12.38.Lg
}
\end{minipage}
\end{center}

\large

Quantum Chromodynamics (QCD) is asymptotically free and its high energy 
behavior has been tested to one-loop accuracy. On the other hand, the 
low-energy behavior has become a very active research field in the 
past years. Various hadronic resonances act as suitable labs for 
exploring the nonperturbative QCD.  
The inner structure of nucleon and mesons and their interactions is of central
importance in nuclear and particle physics. 
Internationally there are a number of experimental  
collaborations, like TJNAL (former CEBAF), 
COSY, ELSA (Bonn), MAMI (Mainz) and Spring8 (Japan), 
which will extensively study the excitation of higher 
nucleon resonances.

Among various baryon resonances, negative parity resonance $N^\ast (1535)$
is particularly interesting, which dominates the $\eta$ meson photo- or 
electro-production on a nucleon. The branching ratio for the decay 
$N^\ast \to \eta N$ is comparable with that for $N^\ast \to \pi N$. 
Considering the phase space difference and using the experimental decay 
width of $N^\ast$ \cite{review}, we get $g_{\eta N N^\ast}=2$ 
and $g_{\pi N N^\ast}=0.7$. 
The latter value is in strong contrast with the pion nucleon coupling
$g_{\pi NN}=13.4$. Thus arises the question: why is the coupling 
$g_{\pi N N^\ast}$ so small compared with $g_{\pi NN}$?

Whether the coupling $g_{\pi N N^\ast}$ is strongly suppressed is under 
heated debate in literature. In a recent extended coupled channel 
analysis of $\pi N$ scattering, the J$\ddot{u}$lich group used 
$g_{\eta N N^\ast}=1.94$ and $g_{\pi N N^\ast}<0.12$ \cite{julich}. 
Jido, Oka and Hosaka suggested in a recent letter that $\pi NN^\ast$ 
coupling is strongly suppressed as a consequence of chiral symmetry \cite{jido}. 
Their argument was based on a pion-nucleon correlator of two baryon 
interpolating fields. The chiral transformation properties of their 
interpolating fields then imply that the correlator is purely proportional 
to the tensor structure $\gamma_5$, with no piece of the form 
${\hat p}\gamma_5$, which is the relevant structure for $\pi N N^\ast$ coupling.
Based on the above observation they claimed that the coupling $\pi N N^\ast$
vanishes. The above argument was criticized by Birse \cite{birse}. 
He pointed out that the ${\hat p}\gamma_5$ piece of the correlator is a sum 
of all possible pion-baryon couplings that can contribute. Hence the absence
of a ${\hat p}\gamma_5$ piece is a statement about the particular combination
of the pion-baryon coupling and the subtraction terms that correspond to the 
chosen interpolating fields. It does not imply that the physical $\pi NN^\ast$
coupling is suppressed. 
With an interpolating field with covariant derivative 
for the $N^\ast$ Kim and Lee used QCD sum rules to estimate 
$g_{\pi NN^\ast} \sim 1.5$ \cite{lee}. 
But in their analysis the continuum and excited states contribution 
is poorly subtracted and the numerical results depend strongly 
on the value of the quark gluon mix condensate 
$\langle {\bar q}g_s \sigma\cdot G q\rangle$, 
which renders their conclusion less convincing.

Although it is widely accepted that QCD is the underlying theory of 
the strong interaction, the self-interaction of the gluons causes 
the infrared behavior and the vacuum of QCD highly nontrivial. 
In the typical hadronic scale QCD is nonperturbative which makes 
the first principle calculation unrealistic except the
lattice QCD approach, which is very computer time consuming. 
So a quantitative calculation of the $\pi NN(1535)$ coupling with 
a tractable and reliable theoretical approach proves valuable.

Among the various nonperturbative methods, QCD sum rules (QSR) is very useful 
in extracting low-lying hadron masses and coupling constants \cite{SVZ}. 
The light cone QCD sum rules (LCQSR)
is quite different from the conventional QSR, which is 
based on the short-distance operator product expansion (OPE). 
The LCQSR is based on the OPE on the light cone, 
which is the expansion over the twists of the operators. The main contribution
comes from the lowest twist operator. Matrix elements of nonlocal operators 
sandwiched between a hadronic state and the vacuum defines the hadron wave
functions. When the LCQSR is used to calculate the coupling constant, the 
double Borel transformation is always invoked so that the excited states and 
the continuum contribution can be treated quite nicely. Moreover, the final 
sum rule depends only on the value of the wave function at a specific point like
$\varphi_{\pi}(u_0 ={1\over 2})$, which is much better known than the whole wave 
function \cite{bely95}.

LCQSR has been used to derive the couplings of pions with heavy mesons 
in full QCD \cite{bely95}, in the limit of $m_Q\to \infty$ 
\cite{zhu1} and $1/m_Q$ correction \cite{zhu3}, the couplings 
of pions with heavy baryons \cite{zhu2}, and various semileptonic
decays of heavy mesons \cite{bagan98} etc.

We shall employ the LCQSR to calculate $\pi NN^\ast$ coupling. 
The continuum and excited states contribution is subtraced 
cleanly within our approach.

We start with the two point function 
\begin{equation}
\Pi (p_1,p_2,q) = \int d^4 x e^{ip x} 
\left \langle 0 \vert {\cal T}
\eta_p (x;s)  {\bar{\eta_n}} (0;t) \vert \pi^+ (q) \right \rangle
\label{eq1}
\end{equation}
with $p_1 =p$, $p_2 = p-q$ and 
the general nucleon interpolating field without derivatives 
which couples to both positive and negative parity nucleon resonances
\begin{equation}
\eta_p (x;s) = \epsilon_{abc} \{
 \left [ u^a (x) {\cal C} d^b (x) \right ] \gamma_5  u^c (x) +
s \left [ u^a (x) {\cal C}\gamma_5 d^b (x) \right ]   u^c (x)
\}
\label{eq2}
\end{equation}
where $a,b,c$ is the color indices, 
${\cal C} = i \gamma_2 \gamma_0$ is the charge conjugation matrix, 
$s$ is the mixing parameter and ${\bar{\eta_p}}={\eta_p}^\dag \gamma_0$.
For the neutron interpolating field, $u \leftrightarrow d$. The Ioffe's
current $\eta_p (x;s=-1)$ couples strongly to the positive parity nucleon 
\cite{IOFFE,espriu}, while it was found that the current $\eta_p (x;s=0.8)$ 
is optimized for negative parity nucleons and couples strongly to 
$N(1535)$ \cite{jido2}.

$\Pi (p_1,p_2,q)$ has the general form
\begin{equation}\label{eq3}
\Pi (p_1,p_2,q) =
 F (p_1 ^2 , p_2 ^2 ,q^2) {\hat q} \gamma_5 +
F_1 (p_1 ^2 , p_2 ^2 ,q^2) \gamma_5 + 
F_2 (p_1 ^2 , p_2 ^2 ,q^2) {\hat p} \gamma_5 +
F_3 (p_1 ^2 , p_2 ^2 ,q^2) \sigma_{\mu\nu}  \gamma_5 p^\mu q^\nu
\end{equation}

It was well known that the sum rules derived from the chiral odd tensor 
structure yield better results than those from the chiral even ones
in the QSR analysis of the nucleon mass and magnetic moment 
\cite{IOFFE,zsl-mag}. Most of the QSR analysis of 
the pion nucleon coupling constant deals with the tensor structure 
${\hat q} \gamma_5$. 
It is important to note that the diagonal transitions like $N\to N$, $N^\ast \to N^\ast$
does not contribute to the tensor structure ${\hat p} \gamma_5$. 
In other words, the function $F_2$ involves solely with the process 
$N^\ast \to N$ and corresponding continuum contribution. 
Based on the above observation we shall focus 
on the chiral odd tensor structure ${\hat p} \gamma_5$ throughout.

The $\pi N N(1535)$ coupling constant $g_{\pi NN^\ast}$ is defined as:
\begin{equation}
\label{def}
{\cal L}_{\pi NN^\ast} =  g_{\pi NN^\ast} {\bar N}
{\bf \tau \cdot \pi} N + h.c.
\end{equation}

At the phenomenological level the eq.(\ref{eq1}) can be expressed as:
\begin{equation}
\Pi (p_1,p_2,q)  = -(m_N +m_{N^\ast})   g_{\pi NN^\ast}
\{  {\lambda_N (s)\lambda_{N^\ast} (t)\over{ (p_1 ^2 - M_N^2) (p_2 ^2 - M_{N^\ast}^2) }} 
-{\lambda_{N^\ast} (s)\lambda_N (t)\over{ (p_1 ^2 - M_{N^\ast}^2) (p_2 ^2 - M_N^2) }} 
\} i {\hat p} \gamma_5 +\cdots
\label{phen}
\end{equation}
where we write the structure ${\hat p} \gamma_5$ explicitly only
and the continuum contribution is denoted by the ellipse. 
$\lambda_N (s)$ is the overlapping amplitude of 
the interpolating current $\eta_N (x)$ with the nucleon state
\begin{equation}
\left \langle 0 \vert \eta (0;s) \vert N (p) \right \rangle 
= \lambda_N(s)  u_N (p)
\label{eq10}
\end{equation}

By the operator expansion on the light-cone
the matrix element of the nonlocal operators between the vacuum and 
pion state defines the two particle pion wave function (PWF). 
Up to twist four the Dirac components of this wave function can be 
written as \cite{bely95}:
\begin{eqnarray}\label{phipi}
<\pi(q)| {\bar d} (x) \gamma_{\mu} \gamma_5 u(0) |0>&=&-i f_{\pi} q_{\mu} 
\int_0^1 du \; e^{iuqx} (\varphi_{\pi}(u) +x^2 g_1(u) + {\cal O}(x^4) ) 
\nonumber \\
&+& f_\pi \big( x_\mu - {x^2 q_\mu \over q x} \big) 
\int_0^1 du \; e^{iuqx}  g_2(u) \hskip 3 pt  , \label{ax} \\
<\pi(q)| {\bar d} (x) i \gamma_5 u(0) |0> &=& {f_{\pi} m_{\pi}^2 \over m_u+m_d} 
\int_0^1 du \; e^{iuqx} \varphi_P(u)  \hskip 3 pt ,
 \label{phip}  \\
<\pi(q)| {\bar d} (x) \sigma_{\mu \nu} \gamma_5 u(0) |0> &=&i(q_\mu x_\nu-q_\nu 
x_\mu)  {f_{\pi} m_{\pi}^2 \over 6 (m_u+m_d)} 
\int_0^1 du \; e^{iuqx} \varphi_\sigma(u)  \hskip 3 pt .
 \label{psigma}
\end{eqnarray}
\noindent 

\begin{eqnarray}
& &<\pi(q)| {\bar d} (x) \sigma_{\alpha \beta} \gamma_5 g_s 
G_{\mu \nu}(ux)u(0) |0>=
\nonumber \\ &&i f_{3 \pi}[(q_\mu q_\alpha g_{\nu \beta}-q_\nu q_\alpha g_{\mu \beta})
-(q_\mu q_\beta g_{\nu \alpha}-q_\nu q_\beta g_{\mu \alpha})]
\int {\cal D}\alpha_i \; 
\varphi_{3 \pi} (\alpha_i) e^{iqx(\alpha_1+v \alpha_3)} \;\;\; ,
\label{p3pi} 
\end{eqnarray}

\begin{eqnarray}
& &<\pi(q)| {\bar d} (x) \gamma_{\mu} \gamma_5 g_s 
G_{\alpha \beta}(vx)u(0) |0>=
\nonumber \\
&&f_{\pi} \Big[ q_{\beta} \Big( g_{\alpha \mu}-{x_{\alpha}q_{\mu} \over q \cdot 
x} \Big) -q_{\alpha} \Big( g_{\beta \mu}-{x_{\beta}q_{\mu} \over q \cdot x} 
\Big) \Big] \int {\cal{D}} \alpha_i \varphi_{\bot}(\alpha_i) 
e^{iqx(\alpha_1 +v \alpha_3)}\nonumber \\
&&+f_{\pi} {q_{\mu} \over q \cdot x } (q_{\alpha} x_{\beta}-q_{\beta} 
x_{\alpha}) \int {\cal{D}} \alpha_i \varphi_{\|} (\alpha_i) 
e^{iqx(\alpha_1 +v \alpha_3)} \hskip 3 pt  \label{gi} 
\end{eqnarray}
\noindent and
\begin{eqnarray}
& &<\pi(q)| {\bar d} (x) \gamma_{\mu}  g_s \tilde G_{\alpha \beta}(vx)u(0) |0>=
\nonumber \\
&&i f_{\pi} 
\Big[ q_{\beta} \Big( g_{\alpha \mu}-{x_{\alpha}q_{\mu} \over q \cdot 
x} \Big) -q_{\alpha} \Big( g_{\beta \mu}-{x_{\beta}q_{\mu} \over q \cdot x} 
\Big) \Big] \int {\cal{D}} \alpha_i \tilde \varphi_{\bot}(\alpha_i) 
e^{iqx(\alpha_1 +v \alpha_3)}\nonumber \\
&&+i f_{\pi} {q_{\mu} \over q \cdot x } (q_{\alpha} x_{\beta}-q_{\beta} 
x_{\alpha}) \int {\cal{D}} \alpha_i \tilde \varphi_{\|} (\alpha_i) 
e^{iqx(\alpha_1 +v \alpha_3)} \hskip 3 pt . \label{git} 
\end{eqnarray}
\noindent 
The operator $\tilde G_{\alpha \beta}$  is the dual of $G_{\alpha \beta}$:
$\tilde G_{\alpha \beta}= {1\over 2} \epsilon_{\alpha \beta \delta \rho} 
G^{\delta \rho} $; ${\cal{D}} \alpha_i$ is defined as 
${\cal{D}} \alpha_i =d \alpha_1 
d \alpha_2 d \alpha_3 \delta(1-\alpha_1 -\alpha_2 
-\alpha_3)$. 
Due to the choice of the
gauge  $x^\mu A_\mu(x) =0$, the path-ordered gauge factor
$P \exp\big(i g_s \int_0^1 du x^\mu A_\mu(u x) \big)$ has been omitted.
The coefficient in front of the r.h.s. of 
eqs. (\ref{phip}), (\ref{psigma})
can be written in terms of the light quark condensate
$<{\bar u} u>$ using the PCAC relation: 
$\displaystyle \mu_{\pi}= {m_\pi^2 \over m_u+m_d}
=-{2 \over f^2_\pi} <{\bar u} u>$. 

The PWF $\varphi_{\pi}(u)$ is associated with the leading twist two 
operator, $g_1(u)$ and $g_2(u)$ correspond to twist four operators, and $\varphi_P(u)$ and 
$\varphi_\sigma (u)$ to twist three ones. 
The function $\varphi_{3 \pi}$ is of twist three, while all the 
PWFs appearing in eqs.(\ref{gi}), (\ref{git}) are of twist four.
The PWFs $\varphi (x_i,\mu)$ ($\mu$ is the renormalization point) 
describe the distribution in longitudinal momenta inside the pion, the 
parameters $x_i$ ($\sum_i x_i=1$) 
representing the fractions of the longitudinal momentum carried 
by the quark, the antiquark and gluon.

The wave function normalizations immediately follow from the definitions
(\ref{phipi})-(\ref{git}):
$\int_0^1 du \; \varphi_\pi(u)=\int_0^1 du \; \varphi_\sigma(u)=1$,
$\int_0^1 du \; g_1(u)={\delta^2/12}$,
$\int {\cal D} \alpha_i \varphi_\bot(\alpha_i)=
\int {\cal D} \alpha_i \varphi_{\|}(\alpha_i)=0$,
$\int {\cal D} \alpha_i \tilde \varphi_\bot(\alpha_i)=-
\int {\cal D} \alpha_i \tilde \varphi_{\|}(\alpha_i)={\delta^2/3}$,
with the parameter $\delta$ defined by 
the matrix element: 
$<\pi(q)| {\bar d} g_s \tilde G_{\alpha \mu} \gamma^\alpha u |0>=
i \delta^2 f_\pi q_\mu$.

The full light quark propagator with both perturbative 
term and contribution from vacuum fields reads
\begin{eqnarray}\label{prop}\nonumber
iS(x)=\langle 0 | T [q(x), {\bar q}(0)] |0\rangle &\\
=i{{\hat x}\over 2\pi^2 x^4} 
-{\langle {\bar q} q\rangle  \over 12}
-{x^2 \over 192}\langle {\bar q}g_s \sigma\cdot G q\rangle &\\ \nonumber
-{ig_s\over 16\pi^2}\int^1_0 du \{
{{\hat x}\over x^2} \sigma\cdot G(ux)-4iu {x_\mu\over x^2} 
G^{\mu\nu}(ux)\gamma_\nu \} +\cdots  & \; ,
\end{eqnarray}
where we have introduced ${\hat x} \equiv x_\mu \gamma^\mu$.

We neglect the four particle component of the pion wave function 	
and express (\ref{eq1}) with the PWFs.
After Fourier transformation and making double Borel transformation 
with the variables $p_1^2$ and $p_2^2$
the single-pole terms in (\ref{phen}) are eliminated and finally we arrive at:
\begin{eqnarray}\label{quark1}\nonumber
(m_N +m_{N^\ast}) [ \lambda_N (s)\lambda_{N^\ast} (t) -
\lambda_{N^\ast} (s)\lambda_N (t)] g_{\pi NN^\ast}
e^{-({ M_N^2\over M_1^2}+{ M_{N^\ast}^2\over M_2^2}) } = &\\ \nonumber
-{f_\pi\over 192\pi^2} (1+s)(1+t)
\varphi_\pi^\prime (u_0) M^6 f_2 ({s_0\over M^2})
+{f_\pi\over 16\pi^2} (1+s)(1+t)
g_1^\prime (u_0) M^4 f_1 ({s_0\over M^2})  &\\ \nonumber
-{f_\pi\over 32\pi^2} [7(1+st)+3(s+t)]
g_2 (u_0) M^4 f_1 ({s_0\over M^2}) 
+{f_\pi\over 288\pi^2} (1+s+t-3st)a\mu_\pi
\varphi_\sigma^\prime (u_0) M^2 f_0 ({s_0\over M^2}) &\\ \nonumber
+{f_\pi\over 4608\pi^2} (1+s)(1+t)\langle g_s^2 G^2\rangle 
\varphi_\pi^\prime (u_0) M^2 f_0 ({s_0\over M^2})
-{f_\pi\over 6912\pi^2} (5+3s+3t-11st)am_0^2\mu_\pi
\varphi_\sigma^\prime (u_0) &\\ \nonumber
+{f_\pi\over 192\pi^2} (s-t)am_0^2\mu_\pi \varphi_P (u_0)
-{f_\pi\over 2304\pi^2} [19(1+st)+7(s+t)] \langle g_s^2 G^2\rangle g_2 (u_0) &\\
-{f_\pi\over 1152\pi^2} (1+s)(1+t)\langle g_s^2 G^2\rangle g_1^\prime (u_0)&\; ,
\end{eqnarray}
where $\mu_{\pi}=1.65$GeV, 
$f_{\pi}=132$MeV, $\langle {\bar q} q \rangle=-(225\mbox{MeV})^3$, 
$\langle {\bar q}g_s\sigma\cdot G q \rangle =m_0^2\langle {\bar q} q \rangle$, 
$m_0^2=0.8$GeV$^2$, $a=-(2\pi )^2 \langle {\bar q}q \rangle$.
$f_n(x)=1-e^{-x}\sum\limits_{k=0}^{n}{x^k\over k!}$ is the factor used 
to subtract the continuum, which is modeled by the 
dispersion integral in the region $s_1 , s_2\ge s_0$, 
$s_0$ is the continuum threshold.
$u_0={M^2_1 \over M^2_1 + M^2_2}$, 
$M^2\equiv {M^2_1M^2_2\over M^2_1+M^2_2}$, 
$M^2_1$, $M^2_2$ are the Borel parameters, 
and $\varphi_\pi^\prime (u_0)  ={d\varphi_\pi (u)\over du}|_{u=u_0}$ etc.

In obtaining (\ref{quark1}) we have used integration by parts 
to absorb the factor $(q\cdot x)$, which leads to the derivatives 
of PWFs in the above formula.
Moreover we have used the double Borel transformation formula:
${{\cal  B}_1}^{M_1^2}_{p_1^2} {{\cal  B}_2}^{M_2^2}_{p_2^2} 
{\Gamma (n)\over [ m^2 -(1-u)p_1^2-up^2_2]^n }=
(M^2)^{2-n} e^{-{m^2\over M^2}} \delta (u-u_0 )$.

We note that the twist four PWFs 
$\varphi_\bot(\alpha_i)$, $ \varphi_{\|}(\alpha_i)$,
$ \tilde \varphi_\bot(\alpha_i)$ and $ \tilde \varphi_{\|}(\alpha_i)$
do not contribute to the chiral odd tensor structures, which was 
first observed in \cite{bely-z}. Moreover the twist three PWF $\varphi_{3 \pi}$
appears in the combination $f_{3\pi} \varphi_{3 \pi} \langle {\bar q} q\rangle q^2$.
In the physical limit $q^2=m_\pi^2 \to 0$, its contribution is negligible, 
which is in contrast with the sum rule for $g_{\pi NN}$ \cite{bely-z}, where 
the twist three PWF $\varphi_{3 \pi}$ plays an important role. 

It is a common practice to adopt the Ioffe current for the ground-state 
nucleon \cite{IOFFE}, i.e., letting $s=-1$. Now the equation (\ref{quark1}) has 
a simple form:
\begin{eqnarray}\label{quark2}\nonumber
(m_N +m_{N^\ast}) [ \lambda_N (s)\lambda_{N^\ast} (t) -
\lambda_{N^\ast} (s)\lambda_N (t)] g_{\pi NN^\ast}
e^{-({ M_N^2\over M_1^2}+{ M_{N^\ast}^2\over M_2^2}) } = &\\ \nonumber
-{f_\pi\over 8\pi^2} (1-t) g_2 (u_0) M^4 f_1 ({s_0\over M^2}) 
+{f_\pi\over 72\pi^2} t a\mu_\pi
\varphi_\sigma^\prime (u_0) M^2 f_0 ({s_0\over M^2}) &\\ \nonumber
-{f_\pi\over 3456\pi^2} (1+7t)am_0^2\mu_\pi
\varphi_\sigma^\prime (u_0) 
-{f_\pi\over 192\pi^2} (1+t)am_0^2\mu_\pi \varphi_P (u_0) &\\
-{f_\pi\over 192\pi^2} (1-t)\langle g_s^2 G^2\rangle g_2 (u_0) & \; ,
\end{eqnarray}
where $t=0.8$.

The various parameters which we adopt are 
$a=0.546\, \mbox{GeV}^3$, 
$\langle g_s^2 G^2\rangle=0.474\, \mbox{GeV}^4$, 
$m_0^2 =0.8\, \mbox{GeV}^2$, $s_0=3.2$GeV$^2$,
$\lambda_N (s=-1)=0.013$GeV$^3$, $\lambda_{N^\ast} (s=0.8)=0.27$GeV$^3$ 
with the formulas in \cite{jido2}. Moreover 
$\lambda_N (s=-1) \gg \lambda_N (t=0.8)$, 
$\lambda_{N^\ast} (t=0.8) \gg \lambda_{N^\ast} (s=-1)$, so it is 
reasonable to discard the term $\lambda_N (t=0.8) \lambda_{N^\ast} (s=-1)$ 
in the eq. (\ref{quark2}). We use the model PWFs presented in \cite{bely95}
to make the numerical analysis.

Our sum rule (\ref{quark2}) is asymmetric with the Borel parameters 
$M_1^2$ and $M_2^2$ due to the significant mass difference of $N$ and $N(1535)$.
It is natural to let $M_1^2=2m_N^2 \beta$, $M_2^2=2m_{N^\ast}^2 \beta$,
where $\beta$ is a scale factor ranging from $1.0$ to $2.0$. In this case 
we have $M^2 =1.28\beta$GeV$^2$, $u_0 =0.27$, $g_2(u_0)=-0.03$GeV$^2$, 
$\varphi_P (u_0)=0.66$ and $\varphi^\prime_\sigma (u_0)=2.63$ at the 
scale $\mu =1$GeV. 

The dependence of $g_{\pi N N^\ast}$ on the continuum threshold and the 
Borel parameter is weak. It is very stable with reasonable variations 
of $s_0$ and $M^2$ as can be seen in FIG 1. Finally we have 
\begin{equation}\label{final}
g_{\pi N N^\ast}=(-) (0.08 \pm 0.03) \; .
\end{equation}

We have included the uncertainty due to the variation of the continuum 
threshold and the Borel parameter $\beta$ in (\ref{final}). 
In other words, only the errors arising from numerical analysis of 
the sum rule (\ref{quark2}) are considered. Other sources of 
uncertainty include: (1) the truncation of OPE on the light cone and keeping 
only the lowest twist operators. For example the four particle 
component of PWF is discarded explicitly; (2) the inherent 
uncertainty due to the detailed shape of PWFs in different models; 
(3) throwing away the term $\lambda_N (t=0.8) \lambda_{N^\ast} (s=-1)$ 
in the eq. (\ref{quark2}); (4) the continuum model etc. 
In the present case the major error comes from the 
uncertainty of PWFs since our final sum rule 
(\ref{quark2}) depends both on the value of PWFs and their derivatives
at $u_0=0.27$. Luckily around $u_0=0.27$ different model PWFs 
have a shape roughly consistent with each other. So a more 
conservative estimate is to enlarge the error by a factor of two.
Now we have 
\begin{equation}\label{final2}
g_{\pi N N^\ast}=(-) (0.08 \pm 0.06) \; .
\end{equation}
We want to point out that one should not be too serious about 
the specific number. What's important is 
the fact that $|g_{\pi N N^\ast}| \ll |g_{\pi N N}|$. 
Our result is consistent with the conclusions in Refs. \cite{julich,jido}.

In summary we have used LCQSR to calculate the $\pi N N^\ast$ coupling. 
The continuum and the excited states contribution is subtracted rather 
cleanly through the double Borel transformation with respect to the 
nucleon and $N(1535)$ momenta, $p_1^2$, $p_2^2=(p-q)^2$.
Our result shows explicitly the suppression of $g_{\pi N N^\ast}$.

\vspace{0.8cm} {\it Acknowledgments:\/} This work  
was supported by the National Natural Science Foundation of China
and the National Postdoctoral Science Foundation of China.
S.-L. Zhu thanks Dr. D. Jido for useful communications.

{\bf Figure Captions}
\vspace{2ex}
\begin{center}
\begin{minipage}{130mm}
{\sf FIG 1.} \small{The sum rule for $g_{\pi NN^\ast}$ as a function of the scale 
parameter $\beta$ with the continuum threshold $s_0=3.4, 3.2, 3.0$ GeV$^2$ 
using the PWFs in \cite{bely95}. }
\end{minipage}
\end{center}
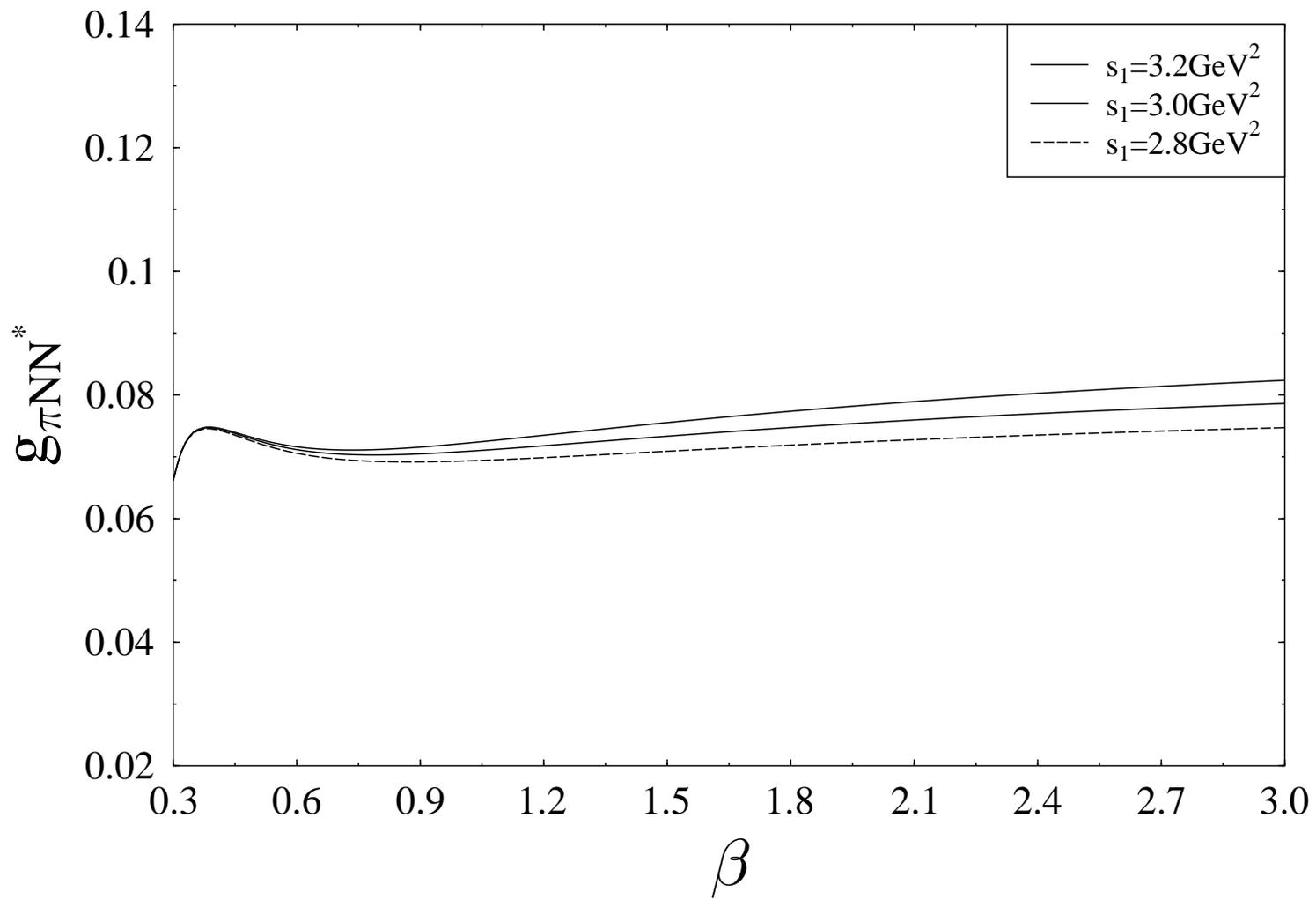
\end{document}